Prediction of the Magnetotoroidic Effect from Atomistic Simulations

Wei Ren and L. Bellaiche

*Physics Department, University of Arkansas, Fayetteville, AR 72701, USA*

An effective Hamiltonian technique is used to investigate the effect of applying *curled electric fields* on physical properties of stress-free $BiFeO_3$ dots being under open-circuit electrical boundary conditions. It is discovered that such fields can lead to a control of not only the magnitude but also the direction of the *magnetization*. On a microscopic point of view, such control originates from the field-induced transformation or switching of electrical vortices and their couplings with oxygen octahedral tilts and magnetic dipoles. This control involves striking intermediate states, and constitutes a novel phenomenon that can be termed as ``magnetotoroidic'' effect.

Multiferroics form an important class of materials mostly because they can exhibit a coupling between magnetic and electric properties that has the potential to lead to the design of novel devices [1, 2]. In these intriguing systems, the magnitude and even the direction of the magnetic order parameter can be modified by applying an *homogeneous* electric field along specific directions [3-6], since such field is the conjugate field of the electrical polarization (this field therefore strongly affects the polarization, that in turn modifies magnetism through direct or indirect couplings between magnetic and electric dipoles in multiferroics). Such desired-control of magnetism by electric fields constitutes the so-called *magnetoelectric* (*ME*) effect.

Interestingly, the ground-state of stress-free zero-dimensional ferroelectrics and multiferroics being under open-circuit-like electrical boundary conditions has been predicted to exhibit an electrical vortex [7, 8]. It therefore does not possess any spontaneous electrical polarization but rather exhibits the so-called electrical toroidal moment as its electric order parameter. Since the conjugate field of the electrical toroidal moment is a *curled* electric field [9-11] rather than an homogeneous electric field, it is legitimate to wonder if the *magnetic* order parameter can be controlled via the application of curled electric fields in multiferroic nanodots being under open-circuit-like boundary conditions. Such hypothetical control will constitute a novel phenomenon that can be termed as the magnetotoroidic (*MT*) effect following the classification of Ref. [12]. Other open questions are: (i) the *microscopic* origin of the *MT* effect (if it exists) and how it is related to the ``normal'' *ME* effect (if any); and (ii) if the hypothetical *MT* effect involves the formation of original intermediate states, based on the fact that the response of electric/magnetic vortices to electric/magnetic field has been documented to induce particularly complex states in low-dimensional ferroelectric/ferromagnetic systems [13, 14].

The aims of this Letter are to address all the aforementioned issues, by using a parameterized effective Hamiltonian for mimicking properties of BiFeO$_3$ (BFO) nanodots being subject to curled electric fields. The magnetotoroidic effect is indeed discovered here for the first time, to the best of our knowledge. Not only the magnitude but also the direction of the magnetization (and antiferromagnetic vector) can be varied when playing with the magnitude and direction of a vector (to be discussed below) quantifying the curled electric field. Our atomistic computations further reveal that the *MT* effect in BiFeO$_3$ nanodots microscopically originates from an interplay between magnetic dipoles, electric vortices and oxygen octahedral tilts. This *MT* effect can involve different regions that correspond to a specific window of applied curled electric fields and that are each associated with striking configurations (such as an electrical antivortex coexisting with a pair of electrical vortices). Each different region has its own linear *MT* coefficient. These different *MT* coefficients can even have opposite sign, while they all are of the same order of magnitude than the linear *ME* effect.

Here, the stress-free BFO nanodots are mostly modeled by a 12×12×12 supercell, hence corresponding to a lateral size of around 5nm. Note that we also carried out simulations with a 16×16×16 supercell and found similar results and conclusions. The mimicked dots are under open-circuit electrical boundary conditions, implying that there is no possible screening of polarization-induced surface charges. Their total internal energy is provided by the effective Hamiltonian approach developed and used in Ref. [8, 15]. In addition to the homogeneous strain tensor, the degrees of freedom of such scheme are the $p_i$ electric dipoles, the oxygen octahedron tilts, magnetic dipoles, and inhomogeneous strains in *each* 5-atom unit cell *i*. Oxygen octahedral tilts in cell *i* are characterized by the antiferrodistortive (AFD) $\omega_i$ vector, whose direction is the axis about which the $FeO_6$ octahedron of cell *i* tilts while its magnitude provides the angle of such tilting. The total internal energy of this effective Hamiltonian approach is used in Monte Carlo simulations that typically run over 50,000 sweeps for equilibrating the system and then over another 50,000 sweeps for obtaining statistical averages of physical properties. Note that effective Hamiltonian approaches (i) correctly predict a R3c ground state in which a spontaneous polarization lies along the [111] pseudo-cubic direction while the oxygen octahedra tilt about this same direction in BFO bulks [16]; (ii) provide accurate Neel and Curie temperatures, and intrinsic magnetoelectric coefficients in BFO bulks [16]; (iii) reproduce the spin-canted magnetic structure that is characterized by a weak magnetization superimposed on a large G-type antiferromagnetic vector in BFO films [17]; and (iv) are even able to quantitatively reproduce the high sensitivity and counterintuitive dependency of the Curie and Neel transition temperatures with epitaxial strains in ultrathin films made of BFO [18].

As indicated by Figs. 1(a) and 1(e), the ground state of the studied stress-free dot having a lateral size of 5 nm adopts a vortex pattern for its electrical dipoles accompanied by a four-domain organization for its oxygen octahedra tiltings. The electric dipolar vortex state consists of four different domains, in which electric dipoles are oriented along [-uu-v], [u-u-v], [u-uv] and [-uuv] pseudo-cubic directions, respectively. The angle between dipoles belonging to neighboring dipolar domains is close to either 71° (for domains alternating along [1-10] in Fig. 1a) or 109° (for domains alternating along [001] in Fig. 1a) [8]. Interestingly, recent density functional theory calculations indicated that 71° and 109° domains give rise to lower energies, in comparison with 180° domains [19]. Furthermore, 71° and 109° ferroelectric domains have been observed experimentally in epitaxial (001) BFO thin films by transmission electron microscopy [20]. Figure 1a is thus consistent with these latter calculations [19] and measurements [20]. The electric vortex state displayed in Fig. 1a is a direct consequence of the open-circuit electrical boundary conditions, and does not possess any spontaneous polarization [7]. It is rather characterized by the so-called electric toroidal moment [7], defined as $\boldsymbol{T} = \frac{1}{2N}\sum_i \boldsymbol{r}_i \times \boldsymbol{p}_i$, where the sum runs over all the $N$ 5-atom cells of the nanodot. In our case, $\boldsymbol{T}$ is lying along the [110] pseudo-cubic direction.

As revealed in Ref.[8], the exotic pattern exhibited by the AFD distortions results from the coupling between oxygen octahedral tilts and the electric vortex. We also numerically further found that the magnetic ground state is characterized by a strong antiferromagnetic vector that is oriented along the [001] pseudo-cubic direction and it coexists with a weak magnetization lying along [-1-10] (i.e. *antiparallel* to the electric toroidal moment). Such magnetic organization allows the antiferromagnetic (AFM) vector and magnetization to be both homogeneous inside the dot and to be perpendicular to each other, as well as, leads to a magnetization which is orthogonal to the $\omega_i$ AFD vectors in any of the four antiferrodistortive domains. This is consistent with the fact that the magnetization in BFO systems arises from a cross-product between the AFM and AFD vectors [8, 21].

Let us now investigate the evolution of physical properties and microstructure for this dot when applying a *curled* electric field [9]. Practically, the simulated curled field is given at a position *r* by $E(r) = \frac{1}{2} S \times r / |r|$, where the vector *S* has a dimension of an electric field and is independent of position. The interaction between this curled field and the local electric dipoles is incorporated by adding $-\sum_i E(r_i) \cdot p_i = -\frac{1}{2} S \cdot \sum_i r_i \times p_i / |r_i|$ to the total internal energy of the effective Hamiltonian approach (where $r_i$ locates the center of cell *i*). This interaction energy hints towards the fact that the electrical toroidal moment will align along *S* for large applied electric fields – which is indeed the case, as we will see below [22].

Figure 2 displays the Cartesian components of four different physical quantities of the studied BFO nanodot at 20K as a function of the magnitude of the *S* vector, when this latter is *antiparallel* to the pseudo-cubic [110] direction (that is opposite to the initial *T*). These quantities are the electric toroidal moment; the averaged AFD-related quantity $\omega$ defined as $\omega = \frac{1}{N} \sum_i \omega_i (-1)^{n_x(i)+n_y(i)+n_z(i)}$ where the sum runs over all the *N* unit cells of the nanodot and where $n_x(i)$, $n_y(i)$ and $n_z(i)$ are the integers locating the 5-atom unit cell *i* centered around $r_i = [n_x(i)\mathbf{x} + n_y(i)\mathbf{y} + n_z(i)\mathbf{z}]a$ (*a* being the predicted 5-atom cubic lattice parameter at 0K, and **x**, **y**, **z** being the unit vectors along the x-, y- and z-axes, respectively); the G-type AFM vector, *L*; and the ferromagnetic (FM) vector, *M*. A particularly striking feature revealed by Fig. 2a is the existence of three different regions. Region I corresponds to the magnitude of *S* ranging between 0 and around 2MV/cm, and is associated with a decrease of the magnitude of the electrical toroidal moment. At the critical field of around 2MV/cm, the system undergoes a first-order transition that leads to the 180-degree *reversal* of *T*. After this critical field and up to a magnitude of *S* around 12MV/cm, that spans Region II, the electrical toroidal moment is thus aligned along the [-1-10] direction, and roughly linearly increases in magnitude. Region III corresponds to larger curled fields. In this latter Region, *T* continues to lie along *S* but increases in magnitude with a much slower rate than in Region II. The reversal of the electrical toroidal moment can be put in use to generate new memory devices with unprecedented storage density [7].

Furthermore, Fig. 2b indicates that the AFD-related quantity $\omega$ also behaves in a different manner in the three aforementioned Regions, but does not change direction (i.e., only its magnitude is varying with *S*). Because of the well-known competition between oxygen octahedral tilting and electric dipoles [23], increasing the magnitude of *S* from zero to 2MV/cm (Region I) results in increasing the magnitude of $\omega$ – since it reduces the magnitude of the electric toroidal moment. The critical field of *T* of 2MV/cm then generates a kink in $\omega$. As further increasing *S* along [-1-10] (Region II) increases the electric toroidal moment along this direction, the AFD motions are disfavored correspondingly and the magnitude of $\omega$ is reduced linearly. Another kink appears in the $\omega$ versus *S* curve at a field of 12MV/cm at the onset of Region III. Upon further increasing the electric field along [-1-10], the magnitude of the AFD-related quantity linearly decreases with the magnitude of *S*, because of the continuing increase of the magnitude of *T*. More interestingly, while the AFM vector is unaffected by the application of the curled electric field (see Fig. 2c), the magnetization adopts the same qualitative response than $\omega$, as consistent with the fact that the FM vector is known to be inherently related to oxygen octahedra tilts in BFO systems [8, 21]. As a result, the magnitude of the magnetization can be controlled by applying *curled* electric fields in BFO dots, which constitutes a novel magnetoelectric effect! Using Schmid's nomenclature classification [12] we may call this phenomenon the linear "magnetotoroidic" effect.

Three different values of the linear magnetotoroidic (MT) coefficient, a (quantifying the change in the magnitude of the magnetization with the magnitude of *S*), can be extracted from Fig. 2d: one positive and equal to $18.3 \times 10^{-7}$ C/Tm$^2$ in Region I; another negative of $-11.8 \times 10^{-7}$ C/Tm$^2$ in Region II; and another one that is also negative and equal to $-7.5 \times 10^{-7}$ C/Tm$^2$ in Region III. In other words, not only the magnitude of the linear MT coefficient but also its sign can be changed, depending on the curled field! It is interesting to realize that these predicted a coefficients are of the same order of magnitude as the linear magnetoelectric coefficient characterizing the change in magnetization (respectively, polarization) in BFO *systems* under *homogeneous* electric fields (respectively, magnetic fields) – that has been reported to be around $4.1$—$13 \times 10^{-7}$ C/Tm$^2$ [17, 24, 25].

Figures 1 show the corresponding evolution of the patterns for the electric dipoles and AFD distortions. As the curled field gradually increases in Region I, small subvortex nucleation occurs at the four corner positions of the nanodot (see Fig. 1b). These cornered structures generate *opposite* toroidal moments compared with the centered vortex for zero field, thus reducing the total *T*. In Region II, the curled electric field generates a complex state (see Fig. 2c), that consists of two adjacent vortices whose chiralities are identical but opposite to the single vortex of the ground state – which is consistent with the 180 degree reversal of the toroidal moment displayed in Fig. 1a. As shown in Fig. 1c, this pair of vortices is separated by an *antivortex* in the dot center. Note that vortex-antivortex pair is a fundamental process that is well documented in nanoscale magnetization dynamics [26-28] while having just been recently experimentally discovered in ferroelectrics [29]. Moreover, let us recall that any vortex is associated with a winding number of +1 while the corresponding winding number of an antivortex is -1[30,31]. As a result, the total winding number is equal to +1 in Regions II, exactly as for the single vortex state under no field. Finally, when the field magnitude further increases (Region III), the resulting electric dipolar state is a single vortex state, as in the ground state but with an opposite chirality (compare Fig. 1d versus Fig. 1a). Figures 1(f) and 1(g) also reveal that the unusual electric dipolar pattern of Regions I and II lead to novel and complex AFD organizations, because of the coupling between electric dipoles and tilts of oxygen octahedra. The present work therefore not only confirms (for the first time, to the best of our knowledge) the hypothesis of Ref. [12] that a magnetotoroidic effect can indeed exist, but it also provides (for the first time too) the microscopic origins of such unusual effect. Note also that the state evolution summarized in Figs 1(a-d) is substantially different from that in $PbZr_{0.5}Ti_{0.5}O_3$ nanodots under a curled electric field, where a lateral vortex (whose electrical toroidal moment is perpendicular to the original and final vortex axes) exists in the intermediate switching state [9]. Difference in anisotropy (initial electric toroidal moment along [001] in Ref. [9] vs [110] in our case) may play a significant role for the different evolutions.

Let us also investigate an important effect that was not mentioned in Ref.[12], namely if a curled electric field can induce a change of *direction* of the magnetization (rather than ``only'' changes its magnitude, as in Fig. 2d). For that, we now apply a curled field for which *S* is along the pseudo-cubic [-110] direction. Figure 3 reveals that a first-order transition occurs above the critical value of 5.6MV/cm for the magnitude of *S*. Just after this transition, all the studied physical quantities have rotated by 90 degrees from their original directions (except the antiferromagnetic vector that remains unchanged)! For instance, the electrical toroidal moment is along [-110] (i.e., along *S*), *ω* is lying along [-1-10], and the magnetization is parallel to [1-10] (that is, it is antiparallel to *S*). These results thus demonstrate that one can control the *direction* of the magnetization by applying curled electric fields, which constitutes a novel effect. Figure 3a also indicates that approaching the critical field of 5.6MV/cm from below generates a structure that possesses an electrical toroidal moment that is approximately along [010]. This transient vortex leads to switching a single vortex structure with electric toroidal moment aligned along *S* at a critical curled electric field that is reduced by a factor of two with respect to the case for which *S* is along [-1-10] (and that involves the vortex-antivortex state shown in Fig. 2c).

Interestingly, the behaviors of the physical properties displayed in Figs 2 and 3 can be understood by assuming the general coupling rules that are illustrated schematically in Figure 4: (1) *T* and *M* want to be antiparallel (or parallel) to each other; (2) *L* desires to be orthogonal to *M* (and thus also to *T*); and (3) *ω* is perpendicular to *T, M* and *L*. Such rules can also be applied to rotate the direction of the *antiferromagnetic* vector. For instance, we also numerically found (not shown here) that applying curled fields with *S* aligned along [101] will result in the rotation of the AFM vector from [001] to [010] at high fields.

In conclusion, the present computations predict that applying curled electric fields in BFO nanodots can control both magnitude and direction of the magnetization (as well as the direction of the antiferromagnetic vector) because of mutual couplings between electric dipoles, oxygen octahedra tilts and magnetic degrees of freedom. Such control constitutes a novel magnetoelectric effect (termed as magnetotoroidic effect [12]) of potential technological importance. On a microscopic point of view, it also involves the formation of peculiar intermediate states, such as pairs of electrical vortices coexisting with a single antivortex. We therefore hope that the present study enhances the current knowledge of the important fields of multiferroics and nanoscience.

Acknowledgements

This work is supported by DOE grant DE-SC0002220, ONR Grants N00014-04-1-0413, N00014-08-1-0915 and N00014-07-1-0825, NSF grants DMR 0701558 and DMR-0080054. Some computations were made possible thanks to the HPCMO of the DoD and the MRI grant 0722625.

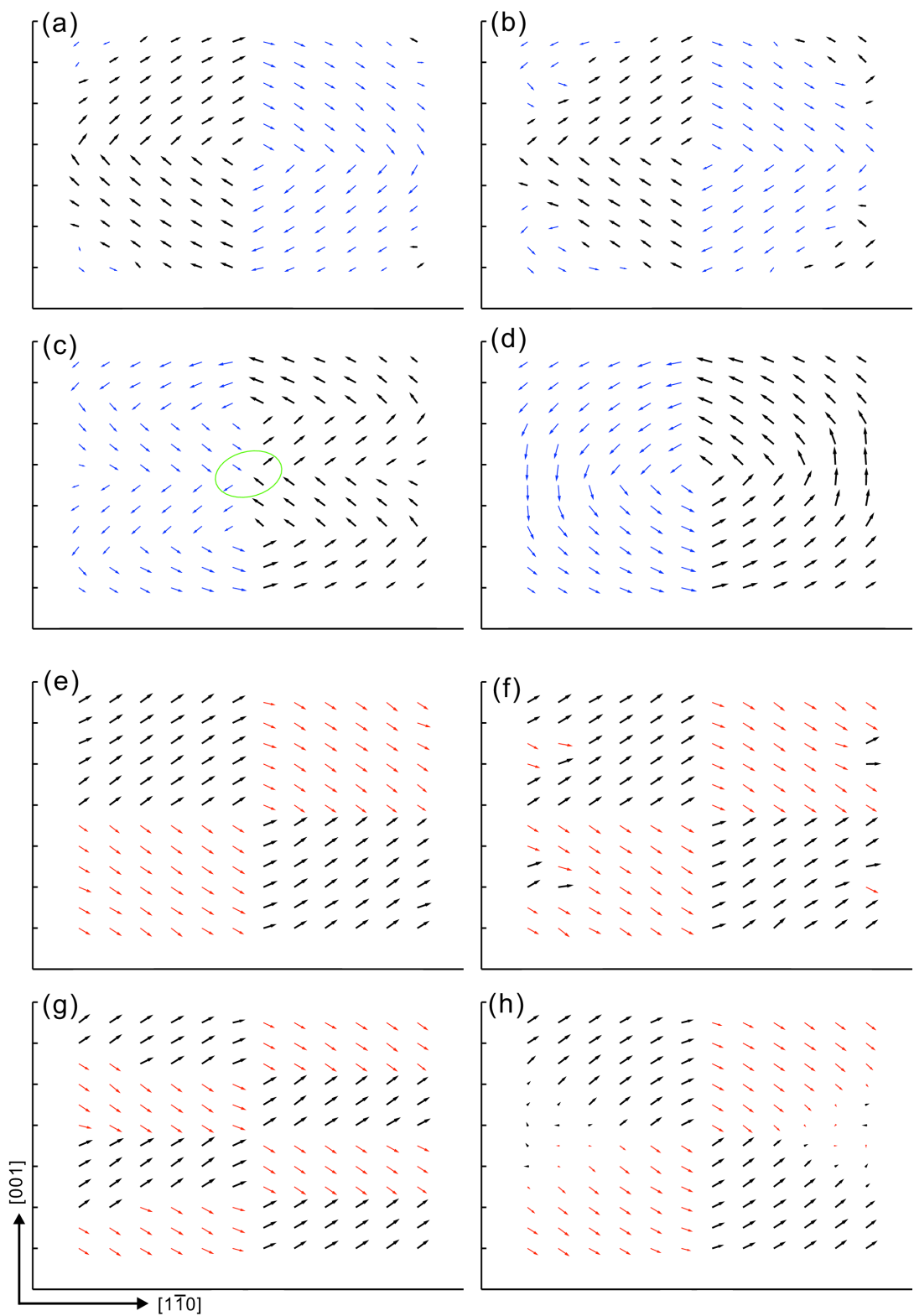

Figure 1 (color online). Snapshots of electric dipolar configurations (Panels a-d) and $(-1)^{nx(i)+ny(i)+nz(i)}\omega_i$ AFD patterns (Panels e-h) in a BFO nanodot of 5 nm lateral size under a curled electric field oriented along [-1-10]. Only a single (110) plane is shown for clarity. Panels (a) and (e) correspond to zero field, while Panels (b) and (f) shows results for a ***S*** vector of 2 MV/cm magnitude (Region I); Panels (c) and (g) show similar results but for a field magnitude of 4 MV/cm (Region II); Finally, Panels (d) and (h) report the predictions for a field magnitude of 12 MV/cm (Region III). The electrical antivortex is highlighted in Panel (c).

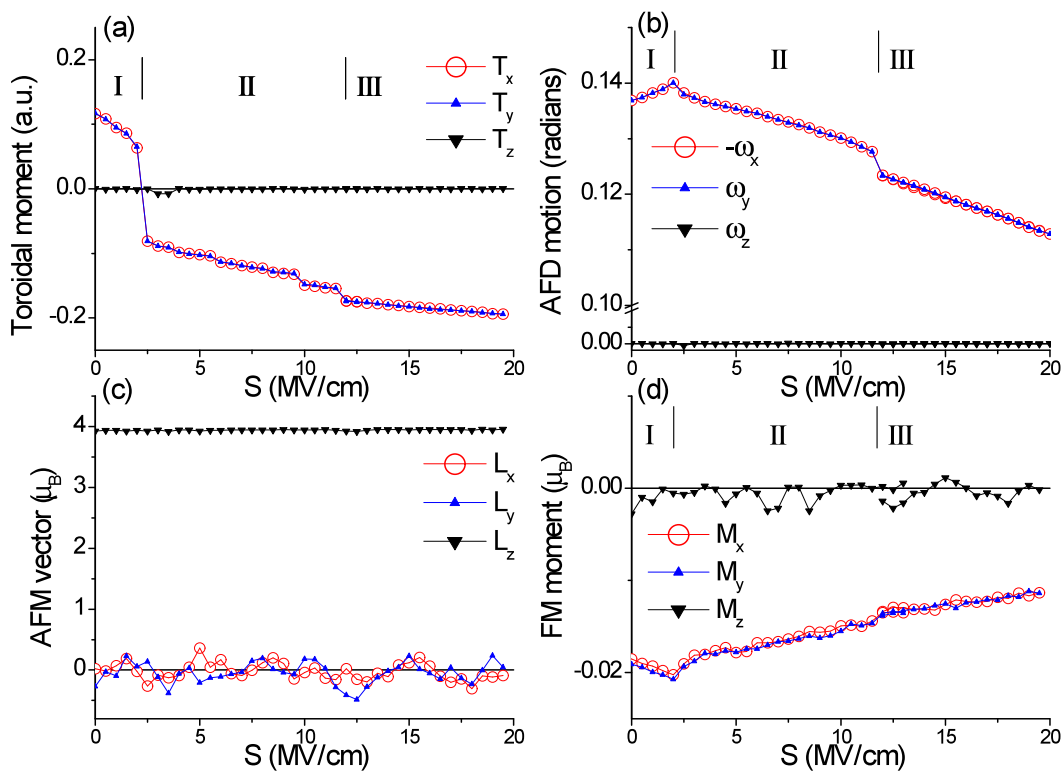

Figure 2 (color online). Electrical toroidal moment (Panel a), AFD-related quantity $\boldsymbol{\omega}$ (Panel b), AFM vector (Panel c) and magnetization (Panel d) as functions of the magnitude of the $\boldsymbol{S}$ vector applied along the [-1-10] direction.

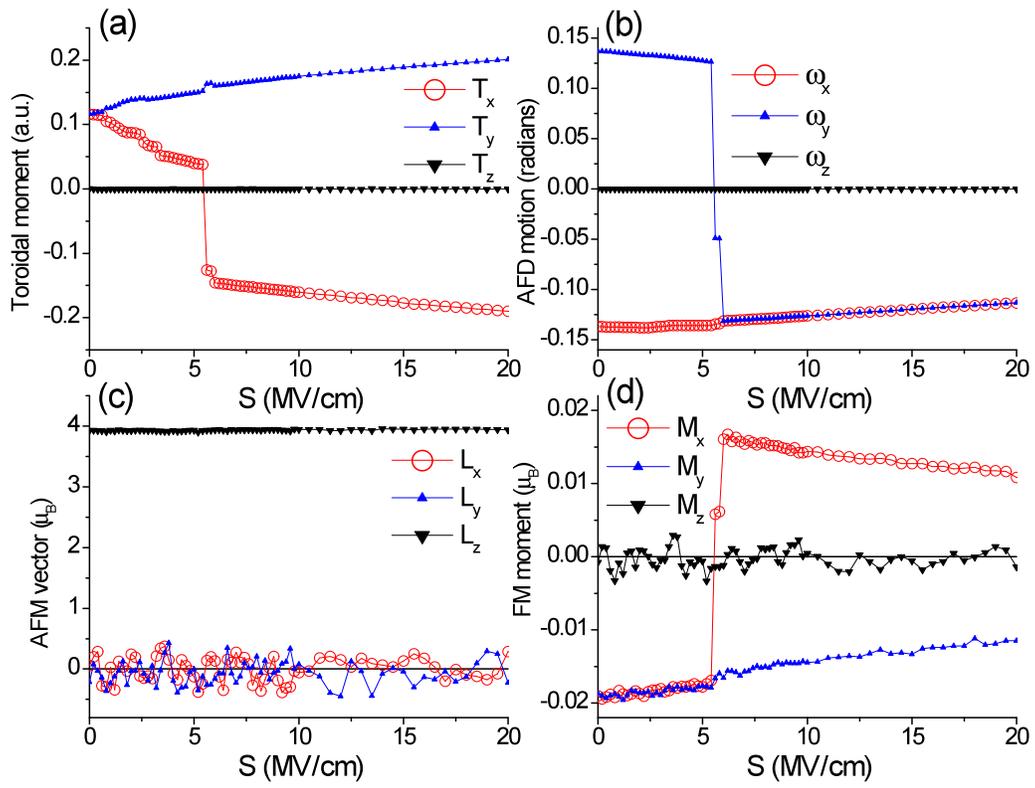

Figure 3 (color online). Same as Figure 2 but for an *S* vector applied along the [-110] direction.

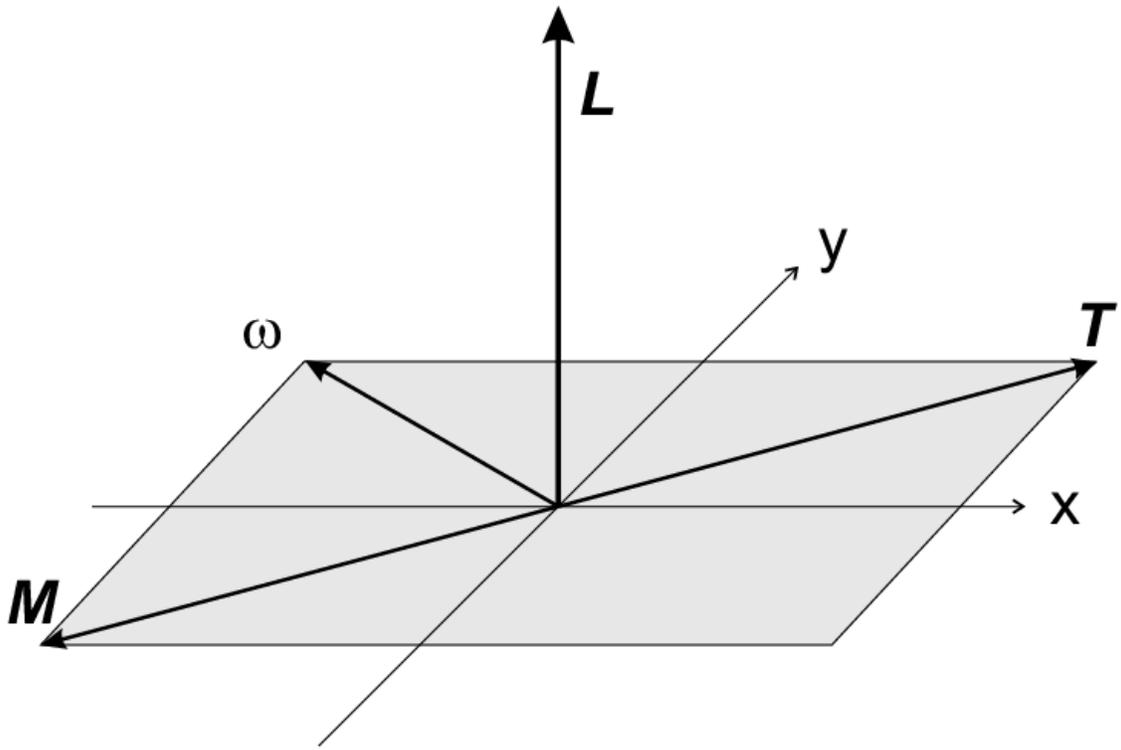

Figure 4. Schematic relationships between the electrical toroidal moment, AFD-related quantity *ω*, magnetization and AFM vector in the ground state of BFO nanodots under open-circuit boundary conditions (not to scale).